\documentclass[prb,twocolumn,showpacs]{revtex4}
\usepackage{bm}
\usepackage{graphicx}
\usepackage{upgreek}
\usepackage{amsmath}
\usepackage{amssymb}

\begin{document}

\title{Classical ratchet effects in heterostructures with a lateral periodic potential}

\author{P. Olbrich,$^1$ J. Karch,$^1$ E.~L. Ivchenko,$^2$ J. Kamann,$^1$ B. M\"arz,$^1$ M. Fehrenbacher,$^1$
D. Weiss,$^1$ and S.~D. Ganichev$^{1}
\footnote{e-mail: sergey.ganichev@physik.uni-regensburg.de}$}
\affiliation{$^1$Terahertz Center, University of Regensburg,
93040 Regensburg, Germany}
\affiliation{$^2$A.~F. Ioffe Physico-Technical Institute, Russian
Academy of Sciences, 194021 St.~Petersburg, Russia}
\begin{abstract}
We study terahertz radiation induced
ratchet currents in 
low dimensional semiconductor structures with a superimposed
one-dimensional lateral periodic potential.
The periodic potential is produced 
by etching a grating into the sample surface
or depositing metal stripes periodically on the sample top. 
Microscopically, the photocurrent generation is based on 
the combined action of the lateral periodic potential,
verified by transport measurements,
and the in-plane modulated pumping
caused by the lateral superlattice. 
We show that a substantial  part of the total current is caused by the polarization-independent Seebeck ratchet effect.
In addition, polarization-dependent photocurrents occur,
which we interpret in terms of their underlying microscopical mechanisms.
As a result, the class of ratchet systems needs to be extended by linear and circular ratchets,
sensitive to linear and circular polarizations of the driving electro-magnetic force.
\end{abstract}
%
\date{\today}
\maketitle

\section{Introduction}
Nonequilibrium spatially-periodic noncentrosymmetric systems are
able to transport particles in the absence of an average
macroscopic force. The directed transport in such systems,
generally known as ratchet effect, has a long history and is
relevant for different fields of
physics.~\cite{reimann,applphys,nori,kotthaus,grifoni,science,samuelson,grifoni2,costache2010}
If this effect is induced by electro-magnetic radiation,
it is usually referred to as photogalvanic (or sometimes photovoltaic) effect,
particularly if breaking of spatial inversion symmetry is
related to the microscopic structure of the
system.~\cite{1,2,chepel,3,regensburg}
Blanter and B\"uttiker~\cite{buttiker2} have shown that one of the
possible realizations of a ratchet is a superlattice (SL)
irradiated by light through a mask of the same period but phase
shifted with respect to the SL yielding a directed current due to local 
electron gas heating. Recently, we have reported an experimental realization 
of this idea with some modifications.~\cite{ratchetPRL}
The photocurrent has been observed in semiconductor
heterostructures with a one-dimensional lateral periodic potential
induced by etching a noncentrosymmetric grating into the sample
cap layer. Hence, the in-plane modulation of the pump radiation appears
not via a mask with periodic structures but due to
near-field effects of terahertz (THz) radiation propagating through the grating.
This photothermal effect, called also Seebeck ratchet effect,~\cite{reimann}
is polarization independent and can be generated even at normal incidence of light. 

Here, we report on the observation and study of radiation induced ratchet effects
sensitive to the plane of polarization of linearly polarized 
light and, in the case of circularly polarized light, 
to the photon helicity. 
The theoretical analysis has enabled us to propose microscopic mechanisms of the
observed circular and linear ratchet effects, and to demonstrate that they  are 
related to the combined action of an out-of-phase periodic potential and an in-plane 
modulated pumping of the two-dimensional electron system (2DES). 
The investigation of these ratchet effects has also been performed on a new set of laterally structured samples with 
a better controlled asymmetry.

The paper is organized as follows. In the next section, we present the theory
of ratchet effects stemming from the combined action of the lateral periodic potential
and the in-plane pumping by the THz field modulated by the near-field diffraction. 
We formulate the model in terms of the classical 
Boltzmann equation for the electron distribution function,
show the position of this model with respect to other electronic ratchets, 
and propose a model picture to interpret the observed photocurrents.
The symmetry analysis in Sect.~\ref{symmetry_analysis} is followed by solving the 
kinetic equation (Sec.~\ref{expansion}) and deriving equations for the Seebeck 
ratchet current (Sec.~\ref{Seebeck}) and polarization-dependent photocurrents (Sec.~\ref{polarization}). 
In Sec.~\ref{samples}, we describe details of the sample preparation and give a short 
overview of the experimental technique. The experimental results are presented and 
discussed in Sec.~\ref{results}. Section \ref{summary} summarizes the study.

\begin{figure}[t]
\includegraphics[width=0.8\linewidth]{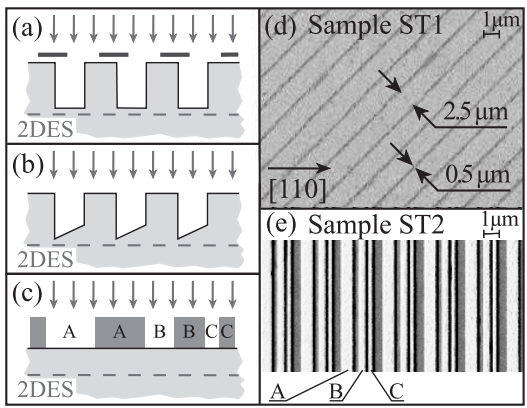}
\caption{Sample design. (a) Blanter and B\"uttiker's geometry. 
(b) The experimental geometry of the first set of samples (type 1, ST1) with an asymmetric groove profile. 
(c) The geometry of the second set of samples (type 2, ST2) with supercells ABCABC... of metallic stripes on top of the sample. 
(d) Electron micrograph of the first set of samples (ST1). 
(e) Electron micrograph of the second set of samples (ST2). 
Here, the widths of the patterns A, B and C are 1~$\upmu$m, 0.6~$\upmu$m and 0.3~$\upmu$m, respectively. 
} \label{figure1}
\end{figure}

\section{Model} \label{model}
We consider a quantum well (QW) structure modulated by a one-dimensional 
periodic lateral potential $V(x)$ with
the period $d$: $V(x + d) = V(x)$. Hereafter we use the right-handed coordinate
system $x,y,z$ with the axes $x,y$ laying in the interface plane and the 
axis $z$ parallel to the growth direction.
In addition to the static potential $V(x)$, the two-dimensional electron gas
is subjected to the action of an in-plane time-dependent electric field 
\[
{\bm E}(x,t) = {\bm E}_{\omega}(x) {\rm e}^{- {\rm i} \omega t} + {\bm
E}_{\omega}^*(x) {\rm e}^{ {\rm i} \omega t}
\]
with the amplitude ${\bm E}_{\omega}(x)$ modulated along the $x$ axis with 
the same period as the static lateral potential: ${\bm E}_{\omega}(x+d) = {\bm E}_{\omega}(x)$. 
The electric field can be linearly polarized with
\[ 
{\rm Im}[E_{\omega, \alpha}(x) E^*_{\omega, \beta}(x)] = 0\hspace{4 mm} (\alpha, \beta = x,y)\:,
\]
or circularly polarized 
with $E_{\omega, y}(x) = \mp {\rm i} E_{\omega, x}(x)$ for $\sigma_+$ 
and $\sigma_-$ polarization, respectively.
Note that the signs in the latter equation correspond to an experimental geometry 
where the light propagates anti-parallel to $z$. 

We will describe the ratchet effects by using the classical Boltzmann equation 
for the electron distribution function
$f_{\bm k}(x,t)$, namely,
\begin{equation} \label{boltzmann}
\left( \frac{\partial}{\partial t} + {\bm v}_{{\bm k},x} \frac{\partial}{\partial x} + \frac{{\bm F}(x,t)}{\hbar}
\frac{\partial}{\partial {\bm k}} \right) f_{\bm k}(x,t) + Q_{\bm k}  = 0\:.
\end{equation}
Here, ${\bm k} = (k_x, k_y)$ and ${\bm v}_{\bm k} = \hbar {\bm k}/m^*$ are the 
two-dimensional electron wave vector and velocity, $m^*$ is the electron effective 
mass, ${\bm F}(x,t)$ is the force consisting of two terms
\begin{equation} \label{force}
{\bm F}(x,t) = - \frac{d V(x)}{d x}\ \hat{\bm e}_x + e {\bm E}(x,t)\:,
\end{equation}
$e$ is the electron charge (negative), ${\hat{\bm e}}_x$ is a unit vector 
along the $x$ axis, and $Q_{\bm k}$ is the collision integral responsible for electron 
momentum and energy relaxation. Equation~(\ref{boltzmann}) is valid for a weak and smooth potential 
satisfying the conditions $|V(x)| \ll \varepsilon_e$ and $q = 2 \pi/d \ll k_e$, where 
$k_e$ is the typical electron wave vector and $\varepsilon_e = \hbar^2 k_e^2/2m^*$ 
is the typical energy being much larger than the photon energy $\hbar \omega$. 
The quantity of central interest is the average electron current
\begin{equation} \label{current_density}
{\bm j} = 2e \sum_{\bm k} {\bm v}_{\bm k} \bar{f}_{\bm k}\:,
\end{equation}
where the factor 2 takes into account the electron spin degeneracy and the bar means 
averaging over the spatial coordinate $x$ and time $t$.
In order to get a nonvanishing directed current, the force ${\bm F}(x,t)$ should be 
asymmetric which means that there exists no coordinate $x_c$ such that 
${\bm F}(x - x_c,t) = {\bm F}(x_c - x,t)$. Note that both the gradient $dV(x)/dx$ 
and the amplitude ${\bm E}_{\omega}(x)$ can possess centers of inversion but, 
in an asymmetric system, these centers must not coincide. 

In terms of the models reviewed and discussed 
in Ref.~[\onlinecite{reimann}], the system under study is 
a pulsating ratchet described in terms of the distribution function and 
Boltzmann 
equation with the collision integral. It is analogous to a Brownian particle 
in two dimensions with 
coordinates $x$, $y$ and mass $m^*$, which is governed
by Newton's equation of motion
\begin{equation} \label{Newton}
m^* \left(\dot{\bm v} + \eta \bm v \right) = - {\bm \nabla} V(x,y,t) + {\bm \zeta}(t)\:.
\end{equation}
Here, the pulsating force $- {\bm \nabla}V(x,y,t) \equiv {\bm F}(x,t)$ 
is given by Eq.~(\ref{force}),
$\eta$ is the viscous friction coefficient and ${\bm \zeta}(t)$ is a 
randomly fluctuating force in the form of a Gaussian white noise of zero mean. 
The system meets the main guiding 
principles of a ratchet: (i) it is periodic both in space and time, (ii) the 
force vanishes after averaging over space and time, (iii) the system is driven 
permanently out of thermal equilibrium and (iv) the force ${\bm F}(x,t)$ is asymmetric.

Two mechanisms, a polarization-independent one and a polarization-dependent one, contribute 
to the current~(\ref{current_density}). 
Here, we present a qualitative interpretation of these mechanisms based on the 
(static) Ohm's law
\begin{equation} \label{Ohm}
{\bm j} = \frac{e^2 \tau N_0}{m^*} {\bm E}\:,
\end{equation}
with momentum relaxation time $\tau$ and electron density $N_0$, 
and reveal the basic physics behind the ratchet effects under study. 
A more detailed discussion is given in Sects. \ref{Seebeck} and \ref{polarization}. 
In line with the first mechanism, the modulated light field heats the electron gas and causes 
a periodic modulation of the effective electron temperature $\Theta(x)$ which, 
in its turn, leads to a redistribution of the electron density $N(x)$ and appearance 
of an electric-field-induced static correction $\delta N(x) \propto |{\bm E}_{\omega}(x)|^2$. 
The polarization-independent dc current is obtained from Eq.~(\ref{Ohm}) 
if the density $N_0$ and the product $eE$
get replaced by $\delta N(x)$ and $-dV(x)/dx$, respectively, and the current is 
averaged over $x$ as follows
\begin{equation} \label{ratchet1}
j_x = \frac{e \tau}{m^*} \overline{ \left[ \delta N(x) \left( - \frac{dV(x)}{dx} \right)\right]}
= \mu_e \overline{ \left( \delta N(x) \frac{dV(x)}{dx} \right)}\:,
\end{equation}
where $\mu_e$ is the electron mobility $|e|\tau/m^*$. Due to the 
asymmetry of the system, the average
of the product $|{\bm E}_{\omega}(x)|^2[dV(x)/dx]$ is nonzero. In the simplest case where 
\begin{eqnarray} \label{phiEV}
{\bm E}_{\omega}(x) &=& {\bm E}_0 [ 1 + h_1 \cos{(q x + \varphi_E)}] \:,\\ 
V(x) &=& V_1 \cos{(q x + \varphi_V)}\:, \nonumber
\end{eqnarray}
with $h_1$ being real, one has
\begin{eqnarray} \label{averageVE}
j_x &\propto& \overline{|{\bm E}_{\omega}(x)|^2[dV(x)/dx]}\\ &=& |{\bm E}_0|^2 q V_1 h_1  \sin{(\varphi_E - \varphi_V)}\:.
\nonumber
\end{eqnarray}
In a more general case with
\begin{eqnarray} \label{phiEV2}
{\bm E}_{\omega}(x) &=& {\bm E}_0 [ 1 + \sum\limits_{n=1}^{\infty} h_n \cos{(n q x + \varphi_{E,n})}] \:,\\ 
V(x) &=& \sum\limits_{n=1}^{\infty} V_n \cos{(n q x + \varphi_{V,n})} \nonumber
\end{eqnarray}
the current in $x$-direction is given by
\begin{equation} \label{averageVE2}
j_x \propto |{\bm E}_0|^2 \sum\limits_{n=1}^{\infty} n q V_n h_n \sin{(\varphi_{E,n} - \varphi_{V,n})}\:.
\end{equation}

According to a classification suggested in Ref.~[\onlinecite{reimann}], 
a dc current, resulting from a periodic temperature profile 
induced by an ac driving force,  
represents the Seebeck ratchet effect. 
This photothermal effect has been first considered theoretically by Blanter 
and B\"uttiker~\cite{buttiker2} 
(see also Ref.~[\onlinecite{buttiker}]) 
for a lateral superlattice covered by a periodic mask with a shifted phase 
respective to the SL. 
These authors have shown that irradiation of the SL by light through the mask 
results in a directional current perpendicular to the grating due to local electron gas heating. In the present 
work we assume, in accordance with the modulation-doped QWs used in experiment, 
that $|V(x)| \ll \varepsilon_e$ in contrast to the opposite case discussed in Ref.~[\onlinecite{buttiker2}].
In the experiment described below the mask is 
replaced by a one-dimensional array of grooves etched into the top cap of a 
semiconductor heterostructure or periodically arranged metallic stripes on the sample surface. 
The effect of the periodic structures is two-fold: 
Firstly, they generate a weak one-dimensional periodic potential superimposed upon 
the two-dimensional electron gas and, secondly, they modulate, because of near-field 
diffraction, the electric field amplitude and intensity of the incident light making 
them spatially periodic in the interface plane $(x,y)$. The system asymmetry determined 
by the difference between $\varphi_{E,n}$ and $\varphi_{V,n}$ is a natural consequence 
of the asymmetric shape of the grooves, displayed in Fig.~1(b), or of a periodically repeated asymmetric supercell ABCABC... 
where the stripe width follows a 10:6:3 ratio, as displayed in Fig.~1(c). 
The lateral pattern gives rise to spatial modulation of the 
potential $V(x)$ and the near-field amplitude ${\bm E}_{\omega}(x)$  but, because 
of the asymmetry of the pattern, the modulation phases $\varphi_{E,n}$ and $\varphi_{V,n}$ 
are, in general, different.

The polarization-dependent direct current stems from the time-dependent electron density oscillation $\delta N(x,t)$
linear in both the electric field $E_x$ and the lateral force $-dV(x)/dx$. 
The direct current is obtained from Eq.~(\ref{Ohm}) after substituting 
\begin{equation} \label{oscillation}
E \to {\bm E}(x,t)\:,\: N_0 \to \delta N(x,t) = \delta N_{\omega}(x) {\rm e}^{- {\rm i} \omega t} + {\rm c.c.}
\end{equation}
and averaging over $x$ and $t$, i.e.,
\begin{eqnarray} \label{ratchet2}
{\bm j} &=& \frac{e^2 \tau}{m^*} \overline{\left[ \delta N(x,t) {\bm E}(x,t) \right]}\\
&=& 2 |e| \mu_e {\rm Re}\overline{ \left[ \delta N_{\omega}(x) {\bm E}_{\omega}^*(x) \right]}\:. \nonumber
\end{eqnarray}
For the spatial modulation defined by Eq.~(\ref{phiEV}), we get again 
$j \propto |{\bm E}_0|^2 h_1 V_1 \sin{(\varphi_E - \varphi_V)}$. 
Additionally, as will be shown in Sect.~\ref{polarization}, 
this current also depends on the orientation of the polarization plane 
of linearly-polarized light and on the helicity (degree of circular 
polarization) in the case of circularly-polarized light. 
In this connection, we have supplemented
the classification scheme introduced by Reimann~\cite{reimann} by adding {\it linear} 
and {\it circular ratchet} effects in our previous short article.~\cite{ratchetPRL}

\section{Symmetry analysis} \label{symmetry_analysis}
In this section, we will analyze symmetry restrictions imposed on the 
polarization dependence of the ratchet currents.
The system described by Eqs.~(\ref{boltzmann}) or (\ref{Newton}) has the 
point group symmetry C$_s$ consisting of the identity element and the 
reflection $\sigma$ in the plane perpendicular to the $y$ axis. It follows 
then that the current density components $j_x, j_y$ are related to components 
of the polarization unit vector ${\bm e} = {\bm E}_0/|{\bm E}_0|$ by four 
linearly independent coefficients
\begin{eqnarray} \label{2}
j_{x} &=& \bar{I} [\chi_1 + \chi_2 (|e_{x}|^2 - |e_{y}|^2)]\:,\\
j_{y} &=& \bar{I} [ \chi_3  ( e_{x} e^*_{y} + e_{y} e^*_{x} ) - \gamma P_{\rm circ}\hat{e}_z]\:,\nonumber
\end{eqnarray}
where $P_{\rm circ}\hat{\bm e} = {\rm i} ({\bm e} \times {\bm e}^*)$,
$\bar{I}$ is the average light intensity defined by 
$$ 
\bar{I} = \frac{c n_{\omega}}{2 \pi} (|E_{0x}|^2 + |E_{0y}|^2)\:,
$$
$c$ is the light velocity in vacuum, and $n_{\omega}$ is the refractive index. 
Note that in the described geometry the light propagation 
direction and the $z$ axis are anti-parallel causing the minus sign in the second Eq.~(\ref{2}).
The Seebeck ratchet effect is connected to the coefficient $\chi_1$, while the remaining 
three coefficients describe the linear ($\chi_2, \chi_3$) and circular ($\gamma$) ratchet effects. 

Equations~(\ref{2}) should be compared to the ones of corresponding unpatterned samples,
called reference samples below,
or structures with a symmetric potential. 
One-sided modulation-doped QWs, grown along the crystallographic [001] direction 
of zinc-blende-lattice semiconductors have point-group symmetry C$_{2 v}$
which excludes in-plane currents for normal incidence where $E_{0z} = 0$, 
in contrast to the ratchet currents (\ref{2}), allowed for this geometry. Under oblique 
incidence, the reference samples admit directional photogalvanic electric currents
perpendicular to the plane of incidence~\cite{physicaE2002,10}
\begin{eqnarray} \label{pgec2v}
&&j_{x'} = I [ \chi_{x'x'z} \left( e_{x'} e_z^* + e_z e_{x'}^* \right) + \gamma_{x'y'} P_{\rm circ}\hat{e}_{y'}]\:,\\
&&j_{y'} = I [ \chi_{y'y'z} \left( e_{y'} e_z^* + e_z e_{y'}^* \right) + \gamma_{y'x'} P_{\rm circ}\hat{e}_{x'}] \:, \nonumber
\end{eqnarray}
which are caused by the lack of an inversion center in the reference samples at the atomic level. 
Here, $x'$ and $y'$ denote the axes [$\bar{1}10$] and [110], respectively, 
${\bm \chi}$ and ${\bm \gamma}$ are a third-order tensor and a
second-order pseudotensor describing the linear (LPGE) and circular (CPGE) photogalvanic effects, respectively. 
Equations~(\ref{pgec2v}) show that in reference samples a photocurrent  
can be generated only at oblique incidence ($z$-component of the radiation electric field is needed). 
This is in contrast to the asymmetric lateral structures where the current given by Eqs.~(\ref{2}) reaches
a maximum at normal incidence.

In the lateral structure ST1 sketched in Fig.~1(b), the grooves are oriented along the $[100]$
direction and the axes $x \parallel [100], y \parallel [010]$ in Eqs.~(\ref{2})
are rotated around $z$ by 45$^{\circ}$ with respect to $x', y'$; 
in the sample ST2 of Fig.~1(c) the axes $x,y$ coincide with $x',y'$. 
In both patterned samples ST1 and ST2, obliquely incident light generates both the ratchet current~(\ref{2})
and the photogalvanic current~(\ref{pgec2v}). 
This allows to compare the contributions to the photocurrents due to the lack of inversion symmetry on the atomic
(intrinsic mechanisms) and on the micron scale (periodic grating) experimentally.

\section{Expansion in the perturbation theory} \label{expansion}
The ratchet currents given by Eqs.~(\ref{ratchet1}) and (\ref{ratchet2}) can be obtained by solving 
the kinetic equation
(\ref{boltzmann}) in third order perturbation theory, i.e., 
second order in the electric-field amplitude and first order in the static 
lateral potential. 
In this work, we use the collision integral $Q_{\bm k}$ in the convenient form of a 
sum of the elastic scattering term $Q^{({\rm el.sc.})}_{\bm k}$ and the energy 
relaxation term $Q_{\varepsilon}$. The former is taken in the simplest form
\begin{equation} \label{elastic_term}
Q^{({\rm el.sc.})}_{\bm k} = \frac{f_{\bm k}(x,t) - \langle f_{\bm k}(x,t) \rangle}{\tau}\:,
\end{equation}
where the brackets mean the average over the directions of ${\bm k}$, and 
$\tau$ is the momentum scattering time assumed to be constant.
The term $Q_{\varepsilon}$ is treated in the approximation of effective temperature, see below.
 
The electron distribution function is expanded in powers of the
light electric-field up to the second order,
\begin{equation} \label{expans}
f_{\bm k}({\bm \rho}) = f^{(0)}_{\bm k}(x) + f^{(1)}_{\bm k}(x,t)
+ f^{(2)}_{\bm k}(x,t)\:,
\end{equation}
where $f^{(0)}_{\bm k}(x)$ is the equilibrium distribution function given by
\[
f^{(0)}_{\bm k}(x) = \left[ \exp{ \left( \frac{\varepsilon_{\bm k} +
V(x) - \mu_0}{k_B T}\right)} + 1 \right]^{-1}
\]
with $\mu_0$, $k_B$ and $T$ being the chemical potential,
Boltzmann constant and absolute temperature, respectively, and $\varepsilon_{\bm k}$ being the 
electron energy $\hbar^2k^2/2 m^*$. Here we consider the limit of high temperatures 
and assume that the electron gas obeys a non-degenerate statistics. Then, retaining terms 
of zero and first orders in the lateral potential, we can approximate the equilibrium function by
\begin{equation} \label{equil_distr2}
f^{(0)}_{\bm k}(x) = \left( 1 - \frac{V(x)}{k_B T} \right) {\rm exp}\left( \frac{ \mu_0 - \varepsilon_{\bm k}}{k_B T} \right)\:. 
\end{equation}

The first-order correction is time-dependent and can be arrranged as a sum of two 
complex-conjugate monoharmonic terms
\[
f^{(1)}_{\bm k}(x,t) = {\rm e}^{- {\rm i} \omega t} f^{(1)}_{{\bm
k} \omega} (x) + {\rm e}^{{\rm i} \omega t} f^{(1)*}_{{\bm k}
\omega} (x)\:.
\]
For the second order correction, it is sufficient to retain 
the time-independent contribution $f^{(2)}_{\bm k}(x) \equiv \xi_{\bm k}(x)$ only
and to reduce Eq.~(\ref{current_density}) to
\begin{equation} \label{photocurrent}
{\bm j} = 2 e \sum_{\bm k} {\bm v}_{\bm k}\ \bar{\xi}_{\bm k}\:.
\end{equation}

By successive iteration of the kinetic equation, we obtain equations for 
first- and second-order corrections
\begin{eqnarray} \label{f1equat}
&&\left( - {\rm i} \omega + \frac{1}{\tau} + v_x
\frac{\partial}{\partial x} - \frac{d V(x)}{d x} \frac{1}{\hbar}
\frac{\partial}{\partial k_x} \right) f^{(1)}_{{\bm k}\omega}(x)\\ 
\mbox{} &&\hspace{ 5 mm}-\ \frac{\langle f^{(1)}_{{\bm k} \omega}(x) \rangle}{\tau} + Q^{(1)}_{\varepsilon} = -
\frac{e}{\hbar}\ {\bm E}_{\omega}(x) \frac{\partial}{\partial
{\bm k}} f^{(0)}_{\bm k}(x) \:, \nonumber
\end{eqnarray}
\begin{eqnarray} \label{xi}
\left( \frac{1}{\tau} + v_x \frac{\partial}{\partial x} - \frac{d V(x)}{d x} \frac{1}{\hbar} \frac{\partial
 }{\partial k_x} \right) \xi_{\bm k}(x) - \frac{ \langle \xi_{\bm k}(x) \rangle}{\tau} \\ +~
Q_{\varepsilon}^{(2)}  = - \frac{2 e}{\hbar} {\rm Re} \overline{ \left[ {\bm E}^*_{\omega}(x)
\frac{\partial}{\partial {\bm k}} f^{(1)}_{{\bm k} \omega}(x)
\right]} \:, \nonumber
\end{eqnarray}
where the superscript $j$ in $Q_{\varepsilon}^{(j)}$ labels the order of 
correction to the collision integral.
Next, we multiply terms in the equation for $\xi_{\bm k}(x)$ by 2$e{\bm v}_{\bm k}$, 
sum over ${\bm k}$ and obtain for the ratchet current
\begin{eqnarray}
{\bm j} &=& \frac{2 e \tau}{\hbar} \sum_{\bm k} {\bm v}_{\bm k} \overline{\frac{d V(x)}{d x} \frac{\partial \xi_{\bm k}(x)}{\partial k_x}} \nonumber \\  &-& \frac{4 e^2 \tau}{\hbar} \sum_{\bm k} {\bm v}_{\bm k} {\rm Re} \overline{ \left[ {\bm E}^*_{\omega}(x) \frac{\partial}{\partial {\bm k}} f^{(1)}_{{\bm k} \omega}(x) \right]}\:. \nonumber
\end{eqnarray}
Integrating by parts over ${\bm k}$ and introducing the spatially-modulated 
electron densities
\begin{equation} \label{deltann}
\delta N(x) = 2 \sum_{\bm k} \xi_{\bm k}(x) \hspace{3 mm} \mbox{and} \hspace{3 mm}\delta N_{\omega}(x) = 2 \sum_{\bm k} f^{(1)}_{{\bm k} \omega}(x)\:,
\end{equation}
we arrive at the final equation
\begin{equation} \label{working}
{\bm j} = \mu_e \left\{ \overline{\delta N(x) \frac{d V(x)}{d x} } + 2 |e|{\rm Re}\overline{ [ {\bm E}^*_{\omega}(x)  \delta N_{\omega}(x)]}\right\} \:,
\end{equation}
which is just the sum of the two currents (\ref{ratchet1}) and (\ref{ratchet2}) derived 
heuristically in Sect.~\ref{model}.

The further development of the theory is based on additional assumptions: 
(i) the energy relaxation time $\tau_{\varepsilon}$ is assumed to exceed the momentum 
relaxation time $\tau$, 
(ii) the electron mean free path $l_e = v_T \tau$ and energy diffusion length
$l_{\varepsilon} = v_T \sqrt{\tau \tau_{\varepsilon}}$ (see, e.g., Ref.~[\onlinecite{bass}])
are both small compared with the SL period $d$, 
(iii) we neglect the influence of ac diffusion on the first-order amplitudes 
$f^{(1)}_{{\bm k} \omega}(x)$ which is valid if $v_T q \ll \omega$, where $v_T$ 
is the thermal velocity $\sqrt{2 k_B T/ m^*}$, or, equivalently, if the period of 
the light, $2\pi/\omega$, is shorter than $d/v_T$, the time of the free flight 
of an electron over the spatial period $d$. On the other hand, no restrictions are 
imposed on the value of the product $\omega \tau$.

\section{Seebeck ratchet effect} \label{Seebeck}
To calculate the Seebeck ratchet current (\ref{ratchet1}), represented 
also by the first term in Eq.~(\ref{working}), 
we need to find a static correction $\delta N(x)$ of the spatially modulated electron density. 
Since the derivative $dV(x)/dx$ already enters the right-hand side of Eq.~(\ref{ratchet1}),
this correction can be found neglecting the lateral potential. 
In this case, we can approximately replace the inhomogeneous term in Eq.~(\ref{xi}) by
\begin{eqnarray}
g(\varepsilon_{\bm k}, x) &\equiv& - \frac{2 e}{\hbar} {\rm Re} \left[ {\bm E}^*_{\omega}(x)
\left\langle \frac{\partial}{\partial {\bm k}} f^{(1)}_{{\bm k} \omega }(x)\right\rangle \right] \nonumber\\
&=& \frac{2 e^2 \tau |{\bm E}_{\omega}(x)|^2 }{m^*(1 + \omega^2 \tau^2)}~ 
\frac{\varepsilon_{\bm k} - k_B T}{(k_B T)^2}~ f^{(0)}_{ \bm k}\:,
\nonumber
\end{eqnarray}
where $f^{(0)}_{\bm k} = \exp{[ (\mu_0 - \varepsilon_{\bm k})/k_B T]}$ 
is the equilibrium distribution function normalized to the average electron density 
$N_0$. Equation (\ref{xi}) with the inhomogeneous term $g(\varepsilon_{\bm k},x)$ 
can be reduced to the following macroscopic equations for the two-dimensional 
electron density $N(x)$, local nonequilibrium temperature $\Theta(x)$,
current density $j_x$ and energy flux density ${\cal J}(x)$ 
\begin{subequations} \label{macro}
\begin{eqnarray} 
&&j_x = u \left\{ N(x) \frac{dV(x)}{dx} + \frac{d}{dx}[k_B \Theta(x) N(x)] \right\}, \label{macroa}\\
&&\frac{d j_x}{dx} = 0\:, \label{macrob} \\ \label{macroc}
&& {\cal J} = \left[2 k_B \Theta(x) + V(x)\right] \frac{j_x}{e}
\\ &&\mbox{} \hspace{11 mm} - \frac{2 u}{|e|} N(x) k_B^2 \Theta(x)
\frac{d \Theta(x)}{dx}\:, \nonumber \\ \label{macrod}
&&\frac{d {\cal J}}{d x}  = \hbar \omega G(x) N(x) - \frac{k_B [\Theta(x) -
T]}{\tau_{\varepsilon}} N(x)\:.  
\end{eqnarray}
\end{subequations}
Here, we introduced the energy relaxation time $\tau_{\varepsilon}$ and the 
generation rate $G(x)$ defined as the Drude absorption rate per particle,
\begin{eqnarray} 
G(x) &=& \frac{2 \sum\limits_{\bm k} \varepsilon_{\bm k} g(\varepsilon_{\bm k}, x)}{\hbar \omega N_0} 
=\frac{2 e^2 \tau |{\bm E}_{\omega}(x)|^2 }{m^*\hbar \omega(1 + \omega^2 \tau^2)} \nonumber\\
&=&  \frac{4 \pi e^2}{m^*c n_{\omega}}~ \frac{\tau}{1 + \omega^2 \tau^2}~\frac{I(x)}{\hbar \omega}\:. \nonumber
\end{eqnarray}
For the sake of completeness, we deliberately included into the set of Eqs.~(\ref{macro}) 
terms originating in the lateral potential. One can check that averaging of 
Eq.~(\ref{macroa}) over $x$ leads to Eq.~(\ref{ratchet1}). 

Under homogeneous optical excitation, $G(x) \equiv G_0$, Eqs.~(\ref{macro}) have the
following solution $$k_B \Theta = k_B T + \hbar \omega G_0
\tau_{\varepsilon}\:,\: N(x) = N_0 {\rm e}^{-
V(x)/k_B \Theta}\:,$$ where $N_0$ is $x$-independent. 
For this solution, both
$j_x$ and ${\cal J}$ vanish. 
The current $j_x$ becomes nonzero only if the generation rate $G$ varies spatially. 
For the simple spatial modulation (\ref{phiEV}) of the electric field with a 
small coefficient $h_1$, we write $G(x) = G_0 [1 + 2 h_1 \cos{(q x + \varphi_E)}]$. 
Neglecting the energy diffusion term in Eqs.~(\ref{macro}) we obtain that the 
steady-state
generation produces a stationary periodic electron temperature $\Theta(x)$ 
with $\Theta(x) - \bar{\Theta} \equiv \delta \Theta(x)  =
k_B^{-1} \tau_{\varepsilon} \hbar \omega [G(x) -
G_0]$. Now it follows from Eq.~(\ref{macroa}) that this temperature modulation 
is accompanied by a light-induced periodic correction to the electron density 
$\delta N(x) \approx - N_0 \delta \Theta(x)/ \bar{\Theta}$.

For the lateral potential given by Eq.~(\ref{phiEV}) where the
symmetry of the system is broken by a phase shift between
$V(x)$ and $\Theta(x)$, the final result reads
\begin{eqnarray} \label{chi1}
j_x &=& \chi_1 \bar{I} = \zeta \mu_e N_0 \hbar q \omega \tau_{\varepsilon}  \frac{G_0 V_1}{2k_B T}
\\ 
&=& \zeta \frac{4 \pi e^2 }{\hbar c n_{\omega} } \frac{\hbar q}{m^*}
\frac{\mu_e N_0 \tau \tau_{\varepsilon}}{1 + \omega^2 \tau^2} \frac{\bar{I} V_1}{k_B T}\nonumber
\:,
\end{eqnarray}
where $\bar{I}$ is the averaged light intensity, $\zeta = h_1 \sin{(\varphi_V - \varphi_E)}$ is the
asymmetry parameter related to the inhomogeneous photoexcitation and $\bar{\Theta}$ 
is replaced by $T$. 
The Seebeck ratchet current (\ref{chi1}) is polarization independent and increases 
with decreasing temperature. 
For a more complicated spatial modulation (\ref{phiEV2}), the product $\zeta q V_1$ 
should be replaced by $\sum\limits_{n} n q V_n h_n \sin{(\varphi_{V,n} - \varphi_{E,n})}$.

\section{Polarization-dependent ratchet currents} \label{polarization}
Now we turn to the polarization dependent mechanisms of the currents and discuss the 
linear and circular ratchet effects described in Eqs.~(\ref{2}) by the terms proportional to
$\chi_2, \chi_3$ and $\gamma$. We will show that these ratchet 
currents can also be generated in a lateral SL with the out-of-phase periodic potential $V(x)$ 
and electric field ${\bm E}_{\omega}(x)$. For this purpose we consider the second term 
in Eq.~(\ref{working})
or, equivalently, the contribution (\ref{ratchet2}). The oscillation 
$\delta N_{\omega}(x)$ entering Eqs.~(\ref{ratchet2}), (\ref{working}) and 
defined by Eq.~(\ref{deltann}) satisfies the continuity equation
\begin{equation} \label{contin}
- {\rm i} \omega \delta N_{\omega}(x) + \frac{\partial j_{\omega, x}(x)}{\partial x} = 0\:,
\end{equation}
where $j_{\omega, x}(x)$ is the amplitude of current oscillations at frequency 
$\omega$. It follows from Eq.~(\ref{contin}) that, in order to calculate the current given by Eq.~(\ref{ratchet2}), 
it is sufficient to find a contribution to $\delta N_{\omega}(x)$, linear in the lateral potential 
and a {\it non-modulated} electric field replacing ${\bm E}_{\omega}(x)$ by ${\bm E}_0$ in Eq.~(\ref{f1equat}).

The function $f^{(1)}_{{\bm k} \omega} (x)$ is conveniently rewritten as
\begin{equation}
f^{(1)}_{{\bm k} \omega}(x) = \frac{e {\bm E}_0 {\bm v}_{\bm
k} \tau_{\omega} }{k_B T} f^{(0)}_{\bm k}(x) + F_{{\bm k}\omega}(x)\:,
\end{equation}
where $\tau_{\omega} = \tau/(1 - {\rm i} \omega \tau)$. The correction 
$F_{{\bm k}\omega}(x)$ should be calculated in first order in the lateral 
potential and, in this approximation, satisfies the equation
\begin{eqnarray} \label{Fkomega2}
\left( - {\rm i} \omega + \frac{1}{\tau} + v_x
\frac{\partial}{\partial x} \right) F_{{\bm k} \omega}(x)  -
\frac{ \langle F_{{\bm k} \omega}(x) \rangle}{\tau} + Q_{\varepsilon}^{(1)} \\ = \frac{d
V(x)}{d x} \ \frac{e E_{0 x} \tau_{\omega}}{m^* k_B T}
\ f^{(0)}_{\bm k} \:. \hspace{1 cm} \mbox{} \nonumber
\end{eqnarray}
On summing this equation over ${\bm k}$ and neglecting the ac diffusion, we find
\begin{equation} \label{deltanomega}
\delta N_{\omega}(x) =  \frac{{\rm i} e \tau_{\omega} N_0}{\omega m^* k_B T} \frac{d V(x)}{d x} E_{0x}\:.
\end{equation}
Substitution to Eq.~(\ref{ratchet2}) and averaging over $x$ results in the circular and linear ratchet coefficients
\begin{equation} \label{gammamic}
\gamma = \zeta \frac{\pi e^2 }{\hbar c n_{\omega} } \frac{\hbar q}{m^*}
\frac{V_1}{k_B T} \frac{\mu_e N_0 \tau}{\omega (1 + \omega^2 \tau^2)} \:,
\end{equation}
\[
\chi_2 = \chi_3 = - \omega \tau \gamma\:.
\]
Moreover, the current (\ref{ratchet2}) has a polarization-independent
contribution adding the correction $\delta \chi_1 = - \omega \tau \gamma$ to Eq.~(\ref{chi1}). 
It should be noted that so far the parameter $\tau$ was kept constant. 
In the general case of energy-dependent momentum relaxation time $\tau(\varepsilon)$, 
the ratchet currents acquire an additional contribution proportional to 
$d\ln{\tau(\varepsilon)}/d\ln{\varepsilon}$, see, e.g., Refs.~[\onlinecite{drag,spivak}]. 
We also note that we used the condition $\tau_{\varepsilon} \ll \tau$ while deriving the equations for the coefficients 
$\gamma$ and $\chi_j$. 
If these relaxation time parameters are comparable, the relation between the 
phenomenological coefficients can change but Eqs.~(\ref{chi1}) and (\ref{gammamic}) 
can still be used to estimate the order of magnitude of the currents.

The allowance for polarization-dependent effects is a fundamental difference 
between systems with one and more than one dimensional character of motion. Even if the 
pulsating force is modulated in one dimension, say, along the axis $x$, as 
presented by Eq.~(\ref{force}), but the carriers can move in two dimensions 
$x$ and $y$, both components of the electric field, $E_{\omega,x}$ and $E_{\omega,y}$, 
act on the electron motion. In this case the ratchet current is related not 
only to the squared modulus $|E_{\omega,x}|^2$ as in the one-dimensional ratchet 
with particle's motion along one axis $x$, but also contains contributions 
proportional to $|E_{\omega,y}|^2$ as well as to real and imaginary parts of 
the product $E_{\omega,x} E^*_{\omega,y}$.
The density oscillation (\ref{deltanomega}) is caused by the $x$ component of 
the electric field but, according to Eq.~(\ref{ratchet2}), both components of 
the electric field act on this oscillation resulting in a ratchet current both 
in the $x$ and $y$ directions. Moreover, the current $j_y$ depends on the difference 
between phases of the complex amplitudes $E_{0x}$ and $E_{0y}$. If the phases coincide, 
the light is linearly polarized and gives rise to the linear ratchet current 
proportional to the coefficient $\chi_3$ in Eq.~(\ref{2}). If the phases differ 
by $\pm 90^{\circ}$, the light is circularly polarized and induces the circular 
ratchet current described by the coefficient $\gamma$.

\section{Samples and experimental methods} \label{samples}
We study photocurrents in  GaAs/AlGaAs heterostructures employing two types of
lateral superlattice gratings.

The first type of superlattice (ST1) consists of asymmetrically etched grooves with a
SL period $d$ 
of 2.5~$\upmu$m. A corresponding sketch of the grating and an electronic micrograph are shown in
Figs.~\ref{figure1}(b) and (d), respectively. 
The superlattices were prepared on
molecular-beam epitaxy (001)-grown Si-$\delta$-doped $n$-type
GaAs$/$Al$_{0.25}$Ga$_{0.75}$As QW structures having at 
$T$\,=\,4.2\,K (\,=\,300\,K) a  mobility  $\mu_e \approx 4.8 \times 10^6$\,cm$^{2}$/Vs
($\approx$\,6\,$\times$\,10$^3$\,cm$^{2}$/Vs) and a  carrier density
$N_0$ of 2\,$\times$\,10$^{11}$\,cm$^{-2}$
($\approx$\,1.2\,$\times$\,10$^{11}$\,cm$^{-2}$).
At room temperature the electron mean free path $l_e$ is $0.3\,\upmu$m and, hence,
the condition $l_e \ll d$ holds.
For the experiments we used $5$~mm~$\times$~$5$~mm square shaped samples
oriented along the $[1{\bar 1}0]$- and $[110]$-directions. 
To measure photocurrents, pairs of ohmic contacts were alloyed in the middle
of each sample edge. 
Grooves with 0.5~$\upmu$m width and a period of 2.5~$\upmu$m were obtained by electron beam lithography and subsequent
reactive ion etching using SiCl$_4$. Care was taken not to etch
through the two-dimensional electron system. In order to get a large patterned
area of about 1.4~mm$^2$, 64~squares, each  150~$\upmu$m $\times$ 150~$\upmu$m,
were stitched together. 
In the sample ST1 the one-dimensional grating is oriented 
along the $[100]$ cubic direction, with a slight misalignment of about 4$^{\circ}$. 
The micrographs reveal an asymmetric shape of the grooves: 
the average depth on the right side of a groove is smaller than that
on the left side. The reason for this is ascribed to the
different etching velocities along the [110] and [1$\bar{1}$0]
directions.~\cite{adachi,adachi2} As reference samples for this set of structures we used unpatterned
samples R1 and/or structures R2 with grooves very close to $\left\langle 110\right\rangle$. 
The cross section of these grooves is oriented rather symmetric and does not introduce a structure asymmetry. 

To achieve a better control of the asymmetry (which in the previous set of samples depends on anisotropic etching)
and to enable both transport and photocurrent measurements in one and the same device,
another set of samples, ST2, has been fabricated.
For these devices, the lateral superlattices were prepared on a 
(001)-grown Si-$\delta$-doped
GaAs$/$Al$_{0.28}$Ga$_{0.72}$As heterostructure. 
The room temperature mobility and carrier density in the structures without grating are
$\mu_e = 3.2\times 10^3$~cm$^2$/Vs and $N_0=1.8\times 10^{12}$~cm$^{-2}$.
In order to compare the data of modulated and unmodulated two-dimensional electron systems,
we prepared a Hall bar geometry with a patterned region and an unpatterned reference part, as shown in Fig.~\ref{figST3}.
The SL is defined by
e-beam lithography and deposition of micropatterned gate fingers using
15~nm Ti and 120~nm Au. 
The schematics of the gate fingers, 
consisting of stripes having three different widths 
A = 1~$\upmu$m, B = 0.6~$\upmu$m and C = 0.3~$\upmu$m with ratio A:B:C~=~10:6:3, and
separated by A, B, and C, is shown in Fig.~\ref{figure1}(c) and
a corresponding electron micrograph in Fig.~\ref{figure1}(e).
This asymmetric supercell is  repeated to generate an asymmetric but periodic 
potential superimposed upon the two-dimensional electron system.
The asymmetric supercells ABCABC... are patterned on a 500 $\times$ 140~$\upmu$m$^2$ area
and generate a strain-induced potential in the 2DES with a period $d$ of 3.8~$\upmu$m. 
The gate fingers which are all connected and grounded are oriented along the $\left\langle 110 \right\rangle$ direction, perpendicularly to the Hall bar. 
Both parts of the Hall bar were characterized by magneto-transport measurements
in a top-loading He$^3$/He$^4$ dilution cryostat at 100~mK using standard four-probe lock-in technique. 
To avoid heating of free charge carriers small currents, which do not exceed 100~nA, have been applied. 
The resulting low temperature electron density (mobility) are $N_0=2.3\times 10^{11}$cm$^{-2}$ ($\mu_e=1.1\times 10^6$~cm$^2$/Vs)
in the modulated and $N_0=2.3\times 10^{11}$cm$^{-2}$ ($\mu_e=1.5\times 10^6$~cm$^2$/Vs) in the unmodulated 2DES.

\begin{figure}[t]
\includegraphics[width=0.85\linewidth]{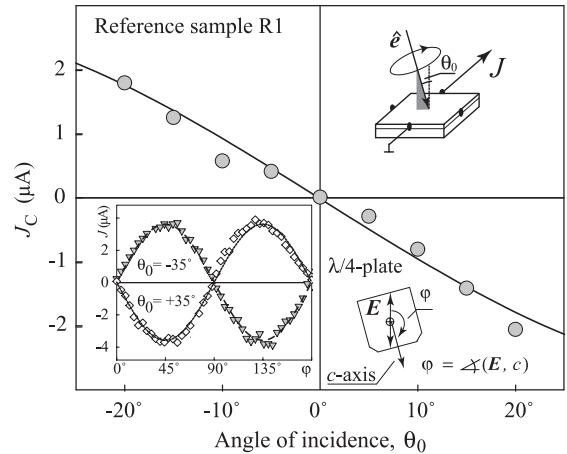}
\caption{Circular photogalvanic current $J_{\rm C} = [J(\varphi = 45^\circ) -
J(\varphi = 135^\circ)]/2$ measured as a function of the angle of incidence
$\theta_0$ in a (001)-oriented GaAs$/$Al$_{0.25}$Ga$_{0.75}$ reference QW 
sample R1 without a lateral structure. 
The current is measured at room temperature in the direction normal to the light propagation. 
The photocurrent is excited by radiation with the wavelength
$\lambda$~=~280~$\upmu$m and power $P \approx 2$~kW. 
The inset (bottom, left) shows the dependence of the total photocurrent $J$ on the
angle $\varphi$ measured for angles of incidence $\theta_0 = \pm 35^\circ$. 
Two other insets (right panels) show,
respectively, the experimental geometry and the quarter-wave plate
which varies the radiation helicity according to $P_{\rm circ}=
\sin 2 \varphi$. 
Full lines are fits to the phenomenological theory
for $C_{2v}$ symmetry relevant for (001)-grown unstructured III-V
QWs and given by Eq.~\protect(\ref{jref}), see Ref.~[\protect\onlinecite{10}]. } \label{fig1}
\end{figure}

\begin{figure}[t]
\includegraphics[width=0.95\linewidth]{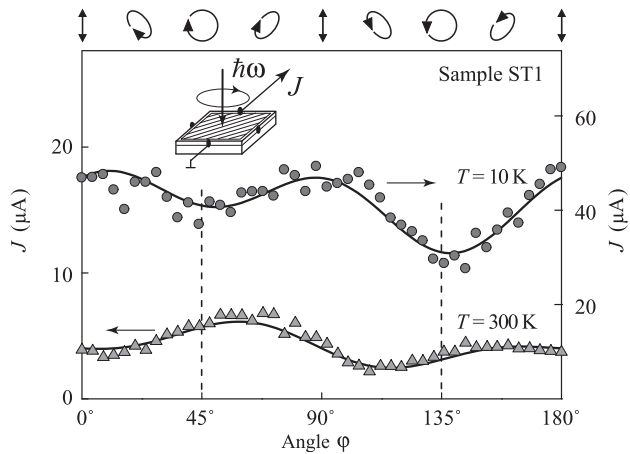}
\caption{Photocurrent measured as a function of the angle $\varphi$ at
normal incidence ($\theta_0 = 0^\circ$) in sample ST1 with the
asymmetric lateral structure prepared along the [100] cubic axis. 
The current is measured at room
temperature and $T = 10$~K, excited by radiation
with the wavelength $\lambda$~=~280~$\upmu$m and power $P \approx 2$~kW.
Full lines are fits to Eq.~\protect(\ref{phenom1}), see also Eq.~\protect(\ref{2}). 
The inset shows the experimental geometry. 
The ellipses on top illustrate the state of polarization for various angles
$\varphi$.} \label{fig2}
\end{figure}

For optical excitation we used a pulsed molecular THz laser with
NH$_{3}$ as an active medium.~\cite{book,JETP1982} Circularly and linearly
polarized radiation pulses of about 100~ns duration with the
wavelength $\lambda$~=~280~$\upmu$m and power $P \simeq 2$~kW were
applied. The photocurrents were induced by indirect intrasubband
(Drude-like) optical transitions in the lowest size-quantized
subband. Various polarization states of the radiation are achieved by
transmitting the linearly polarized ($\bm E \parallel y$) laser beam through $\lambda$/2
or $\lambda$/4 crystal quartz plates. By rotating the $\lambda$/4 plate, one transfers the
linear into elliptical polarization. The polarization states are directly related to the angle
$\varphi$ between the initial linear polarization of the laser light  and the optical axis of the plate, resulting in
$P_{\rm circ} = \sin{2 \varphi}$  for the degree of circular polarization and 
for the bilinear combinations of the polarization vector components in Eqs.~(\ref{2})
\begin{eqnarray} \label{calCS}
&&{\cal S}(\varphi) \equiv e_x e_y^* + e_y e_x^* = \frac12 \sin{4 \varphi}\:, \\
&&{\cal C}(\varphi) \equiv |e_x|^2 - |e_y|^2 = - \cos^2{2 \varphi} \:. \nonumber 
\end{eqnarray} 
If the plane of polarization of linearly polarized light incident upon a $\lambda/2$ plate is 
at an angle $\varphi_{\lambda/2}$ with respect to the slow axis
the plane of polarization of the transmitted light is rotated by an angle 
$\alpha = 2\varphi_{\lambda/2}$ and the above
bilinear combinations are given by
\begin{equation} \label{SCalpha}
{\cal S}(\alpha) =  \sin{2 \alpha}\:,\:{\cal C}(\alpha) = - \cos{2 \alpha} \:.
\end{equation}
Radiation was applied at oblique incidence described by the angle
of incidence $\theta_0$ varying from $-25^\circ$ to +25$^\circ$
(Fig.~\ref{fig1}) and at normal incidence (Fig.~\ref{fig2}).
The current generated by THz light in the unbiased samples was
measured via the voltage drop across a 50~$\Omega$ load resistor
in a closed-circuit configuration. The voltage was recorded with a
storage oscilloscope.

\section{Experimental results} \label{results}
We begin by introducing the results obtained from the reference samples.
In the (001)-oriented unpatterned samples R1, as well as for R2 with 
symmetric groves, a signal is only detectable
under oblique incidence. The photocurrent measured perpendicularly
to the wave vector of the incident light is almost proportional to
the helicity $P_{\rm circ}$ and reverses its direction when the
polarization switches from left- to right-handed circular
(see the inset panel of Fig.~\ref{fig1}). 
In the whole temperature range from
room temperature to 4.2~K 
the variation of the angle of incidence from $\theta_0$ to $-\theta_0$ changes the sign of
the photocurrent $J$. For normal incidence, the photocurrent vanishes. 
This is shown in Fig.~\ref{fig1} where the circular photocurrent in R1 is obtained 
by taking the difference between photoresponses to right- and left-handed radiation 
yielding the CPGE current $J_{C} = [J(\varphi = 45^\circ) - J(\varphi = 135^\circ)]/2$.
Similar results are obtained for the sample R2 (not shown).
The $\theta_0$ and polarization dependencies of the photocurrent 
are in a good agreement with
the phenomenological theory for the circular and linear photogalvanic effects obtained 
for the point group C$_{2v}$.
The total current is well fitted by Eqs.~(\ref{pgec2v}) describing 
the dominating circular photogalvanic current $J_{\rm ref}$ of 
the unpatterned reference samples R1 and R2 as
\begin{equation} \label{jref}
J_{\rm ref} = J_{\rm C, ref} \sin{\theta_0} \xi P_{\rm circ} \:,
\end{equation}
where $J_{\rm 0, ref} = \gamma_{x'y'} t_0^2 I_0$, $I_0$ is the incident intensity, $\xi = t_pt_s/t_0^2$, $t_p$ and
$t_s$ are the Fresnel transmission coefficients for $p$- and
$s$-polarized light, respectively, and $t_0$ is the transmission
coefficient for normal incidence. 
The corresponding fit of  the circular photogalvanic current $J_{\rm ref}$ excited by the
right circularly polarized light ($P_{\rm circ} = 1$)
is shown by the full line in  Fig.~\ref{fig1}. 
A photocurrent, but with substantially smaller magnitude, is also obtained by 
applying linearly polarized radiation. This current is caused by the LPGE and its 
polarization behavior (not shown) is also well described by Eqs.~(\ref{pgec2v}).

\begin{figure}[t]
\includegraphics[width=0.8\linewidth]{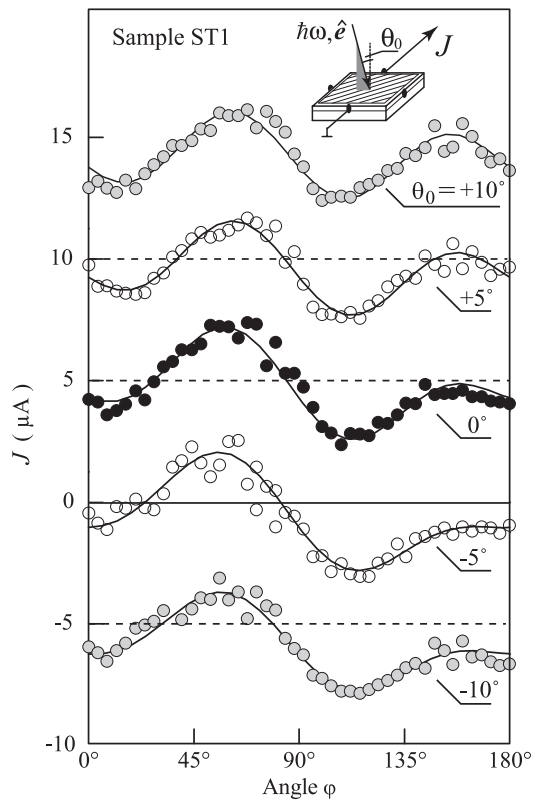}
\caption{Photocurrent measured as a function of the angle $\varphi$ at
various angles of incidence $\theta_0$ in sample ST1 with an
asymmetric lateral structure prepared along the [100] cubic axis. 
The data for $\theta_0 \neq 0$ are shifted by $\pm 5$~$\upmu$A for each $\pm 5^{\circ}$ step in $\theta_0$.
The current is measured at room temperature, excited by radiation
with the wavelength $\lambda$~=~280~$\upmu$m and power $P \approx 2$~kW.
Full lines are fits to Eq.~\protect(\ref{phenom1}) [see also Eq.~\protect(\ref{2})]. 
The inset shows the experimental geometry. 
} \label{fig2wf}
\end{figure}

The situation changes drastically for samples ST1 with asymmetric grating. 
Now a photocurrent can be detected even at normal incidence. The width of the observed
photocurrent pulses is about 100\,ns which corresponds to the duration of the THz
laser pulse. In the patterned samples ST1 where the grooves are oriented along the
$[100]$ direction we have measured a magnitude
of the photocurrent at normal incidence (Fig.~\ref{fig2}) which
is comparable and even larger than the one obtained in the reference
sample R1 at large angles of incidence (Fig.~\ref{fig1}). Moreover,
the polarization behavior has changed.
Figure~\ref{fig2} shows the photocurrent of sample ST1 
as a function of the angle $\varphi$ indicating the helicity.
The current is measured at an angle of 45$^\circ$ with respect to the
axes $x$ and $y$ and can be well fitted by an equal superposition of 
$j_x$ and $j_y$ of Eq.~(\ref{2}) yielding
\begin{equation} \label{phenom1}
J = J_1 + J_2 {\cal C}(\varphi) + J_3 {\cal S}(\varphi) + J_{\rm C} P_{\rm circ}(\varphi)\:.
\end{equation}
Here, the fitting parameters $J_j$ ($j = 1,2,3$) and $J_{\rm C}$ are related to the phenomenological coefficients
$\chi_j$ and $- \gamma$ by the factor $\bar{I}/\sqrt{2}$. 
Figure~\ref{fig2} shows that the helicity dependent photocurrent, denoted as
circular ratchet effect, contributes a substantial fraction to the total current.

\begin{figure}[t]
\includegraphics[width=0.85\linewidth]{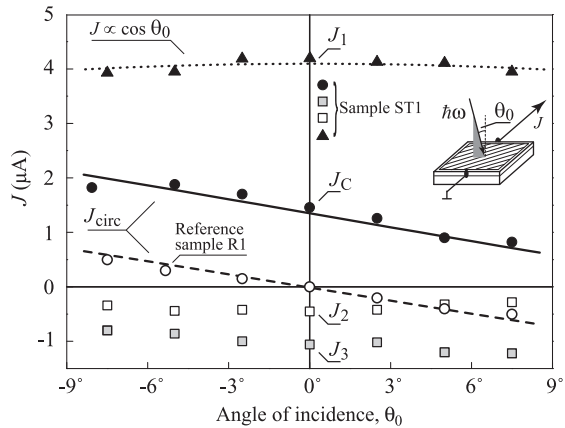}
\caption{Angle of incidence dependence of the photocurrent.
$\bullet$\,and\,$\circ$: Current $J_{\rm C}$ measured in sample ST1 
and the reference sample R1, respectively.
$\blacksquare$, $\square$ and $\blacktriangle$ are
current contributions proportional to $J_3$, $J_2$, and $J_1$.
Dotted line is the fit after Eq.~(\ref{jd}),
solid and dashed lines to Eq.~(\ref{jrefcirc}).
} \label{fig3}
\end{figure}

Equations~(\ref{2}) suggest that the ratchet currents displays a
maximum at normal incidence and is an even function with respect to the angle of incidence.
In order to verify this we measured  the polarization dependence of the photocurrent
for various angles of incidence. The corresponding data are shown in Fig.~\ref{fig2wf}.
In this figure the full lines are fits to Eq.~(\ref{phenom1}) which is applicable not only for normal 
but also for oblique incidence. In the latter case, the current is a sum of contributions 
due to the ratchet effect described by Eqs.~(\ref{2}) and the photogalvanic effect given by Eqs.~(\ref{pgec2v}).
Figure~\ref{fig3} displays the magnitude of the helicity dependent currents as a function of the angle of incidence 
$\theta_0$ for sample ST1 (full circles) and the unstructured reference sample R1 (open circles). 
In addition, the $\theta_0$ dependence of the other three contributions in Eq.~(\ref{phenom1}) 
are shown for the sample ST1.
To extract the 
current $J_{\rm C}$ from the total current we used the fact that the corresponding
contribution to $J$ is proportional to $\sin 2\varphi$ and changes its sign
upon switching the helicity while all the other terms remain unchanged. 
Taking the difference of photocurrents of right- and
left-handed radiation, we get the values of $J_{C}$. 

In the structured sample ST1, the current $J_C$ excited at oblique incidence consists of two contributions. 
The first one has the same origin as the one observed in the reference sample R1 
and is described by Eq.~(\ref{jref}). 
The second one is due to the lateral structure.
The dependence of the circular photocurrent on the angle of incidence
$\theta_0$ can be well fitted by
\begin{equation} \label{jrefcirc}
J_{C} = ( J_{\rm 0, ref} \sin{\theta_0} + J_{\rm C} \cos{\theta_0}) \xi  \:.
\end{equation}

\begin{figure}[t]
\includegraphics[width=0.85\linewidth]{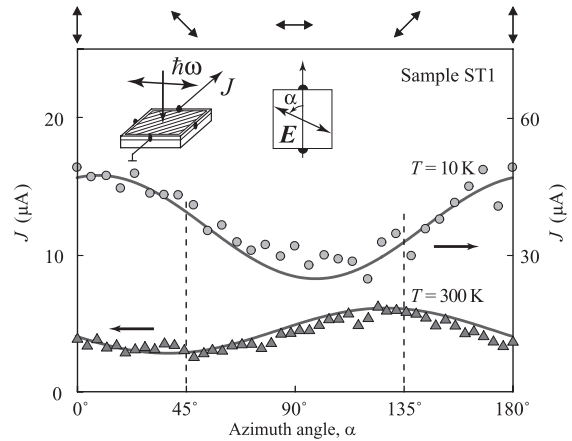}
\caption{Photocurrent $J$ measured as a function of the azimuth angle $\alpha$ 
under normal incidence
at room temperature and $T = 10$~K
in sample ST1 with the asymmetric lateral
structure along the [100] axis.
The photocurrent is excited by linearly polarized radiation with the
wavelength $\lambda$~=~280~$\upmu$m and power $P \approx 2$~kW. 
Full lines are fits to Eq.~\protect(\ref{phenom2}), see also Eq.~\protect(\ref{2}). 
We used for fitting the same values of $J_j$ 
as in the experiments with elliptically polarized radiation, see Fig.~\ref{fig2}. 
Left inset shows the experimental geometry, and right inset
defines the angle $\alpha$. Arrows on top indicate the polarization corresponding 
to various values of
$\protect\alpha$. } \label{fig4}
\end{figure}

Now we turn to the photon helicity independent contributions to
the photocurrent, denoted by the coefficients $J_j$ 
in Eq.~(\ref{phenom1}) which describe photocurrents generated by linearly polarized
radiation. Figure~\ref{fig4} shows the dependence of $J$
on the azimuth angle $\alpha$. We have found that all data can be well fitted by
\begin{equation} \label{phenom2}
J = J_1 + J_2 {\cal C}(\alpha)  + J_3 {\cal S}(\alpha) \:.
\end{equation}
We emphasize that $J_j$ 
are the same fitting parameters as the ones used for the data shown
in Fig.~\ref{fig2}.
Figure~\ref{fig3} shows the dependence of the polarization
independent contribution, proportional to the coefficient $J_1$, on
the  angle of incidence $\theta_0$. 
In this case, the experimental data can be well fitted by
\begin{equation} \label{jd}
J = J_1 \cos{\theta_0} \xi    \:.
\end{equation}
Figures~\ref{fig2} and~\ref{fig4} demonstrate that the dominant
contribution to the photocurrent is polarization independent and
can therefore be obtained by unpolarized radiation.
 
\begin{figure}[b]
\includegraphics[width=0.8\linewidth]{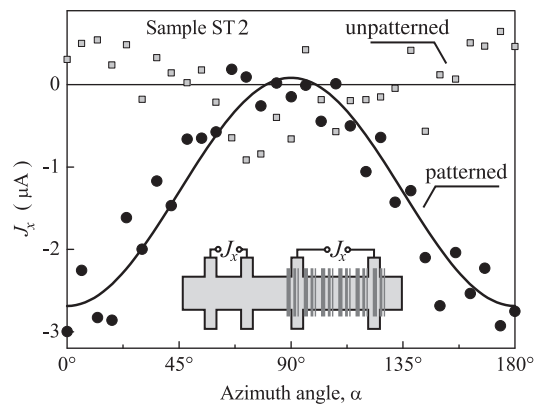}
\caption{Photocurrent $J_x$ measured at sample ST2 as a function of the azimuth angle $\alpha$
at normal incidence and a wavelength of $\lambda$~=~280~$\upmu$m and power $P \approx 9$~kW.
The dependences for both parts with and without asymmetric stripes are shown.
The current induced in the structured part by linearly polarized radiation is well fitted by Eq.~(\ref{phe3}).
The inset displays the design of the Hall bar with the structured and the unpatterned part.
} \label{figST3}
\end{figure}

\begin{figure}[t]
\includegraphics[width=0.8\linewidth]{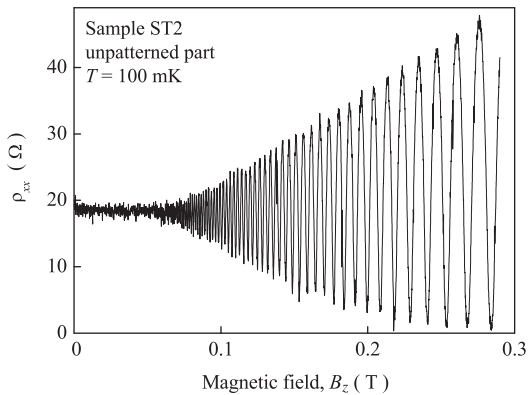}
\caption{Shubnikov-de Haas oscillations measured at $100$~mK in the unpatterned reference section of the ST2 sample.
} \label{transport1}
\end{figure}

The results obtained on the second set of lateral samples, ST2,
are shown in Fig.~\ref{figST3}. 
Again, we start the discussion with the photocurrent in the 
reference, here the unstructured part of the Hall bar (see inset of Fig.~\ref{figST3}).
Also here, a photocurrent is only observed at oblique incidence. This finding
is in agreement with the results above underlining once again that ratchet effects do not
occur in unpatterned structures. 
In the patterned part of the sample, however, a remarkable photocurrent $J_x$
can be observed at normal incidence.
As shown in Fig.~\ref{figST3}, the current, which flows perpendicularly to the asymmetric stripes,
strongly depends on the azimuth angle $\alpha$ of the light's polarization defined above and can be well fitted by
\begin{equation} \label{phe3}
J = J_1 + J_2 {\cal C}(\alpha) \:.
\end{equation}
This is fully in line with the theory of the ratchet effect discussed above (see Secs.~\ref{symmetry_analysis} and \ref{polarization})
and demonstrates that an asymmetric periodic potential
can be controllably introduced by the ABC gate.

\begin{figure}[t]
\includegraphics[width=0.8\linewidth]{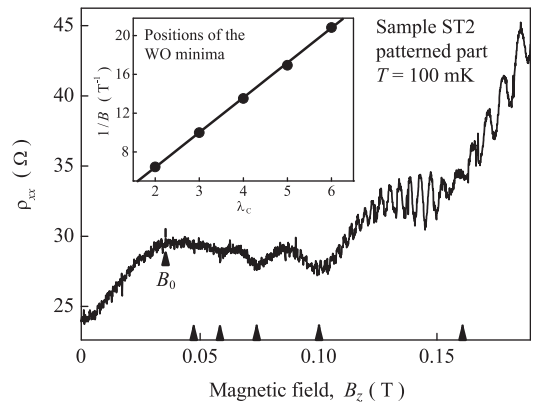}
\caption{Longitudinal resistance $\rho_{xx}$ in the modulated part of the Hall bar sample ST2. 
At low $B$, $1/B$ periodic commensurability oscillation indicate the presence of a weak periodic potential.
The $1/B$ periodicity of the low-field oscillations is evident from the inset
where the oscillation index $\lambda_{\rm C}$ is plotted vs. the resistance minima position $1/B$.
} \label{transport2}
\end{figure}

To check that the 2DES potential is in fact modulated, we carried out magneto-transport measurements at low temperatures.
Corresponding data of the longitudinal resistance  $\rho_{xx}$ of the unpatterned reference area
as a function of the magnetic field $B_z$ are shown in Fig.~\ref{transport1}
and display pronounced Shubnikov de Haas oscillations. 
As the mean free path $l_e$ in the superlattice device is about $9~\upmu$m and hence longer
than the period of the SL as well as much longer than the average distance between neighboring finger strips
we expect commensurability effects to occur.~\cite{Weiss1989}
In this limit the periodic potential causes $1/B$ periodic resistance oscillation where minima are given by the condition
\begin{equation} \label{transport}
2 R_{\rm C} = \left( \lambda_{\rm C} - \frac{1}{4} \right) d \:, \ \lambda_{\rm C} = 1,2,3...
\end{equation}
Here, $2 R_{\rm C}$ is the semi-classical cyclotron orbit diameter and 
$\lambda_{\rm C}$ is the oscillation index. 
Such commensurability (or Weiss-oscillations) are clearly visible at low magnetic fields
of the trace measured in the superlattice part of the sample, as presented in Fig.~\ref{transport2}. 
The SL period $d$ extracted from the Weiss-oscillations (WO) is about $570$~nm
and agrees with one of the Fourier components of the asymmetric periodic potential. 
Why this particular Fourier component dominates is a subject of future investigations. 
Qualitatively it is understandable that the contribution stemming from the full periodicity of $3.8~\upmu$m is cut off
due to scattering, as the circumference of the corresponding cyclotron orbit is about $l_e$,
and that contributions with much smaller periodicity are cut off as the corresponding Fourier coefficients get exponentially damped with increasing surface-2DES distance.~\cite{Gerhardts}
Nonetheless the presence of commensurability effects is a clear signature of the presence of a weak periodic potential. 

\section{Summary} \label{summary}
The lateral grating etched into the sample's surface
or deposition of periodic metal stripes on the sample top
induce a periodic lateral potential acting on the two-dimensional
electron gas. 
As a consequence, the magneto-transport properties of the heterostructure changes
and 1/$B$ oscillations appear at low temperatures in the longitudinal magneto-resistance.
In addition, if illuminated, it modifies the normally-incident
radiation causing its spatial modulation in plane of the electron
gas. If the lateral superlattice is asymmetric the spatial
modulations of the static lateral potential $V(x)$ and the
radiation intensity $I(x)$ are shifted relative to
each other. As a result the product of the static force $-
dV(x)/dx$ and the photothermal modulation of the electron density
$\delta N(x)$ has a non-zero space average and, therefore, a
homogeneous electric current is generated, an effect previously
predicted by Blanter and B\"uttiker.~\cite{buttiker2} The class of 
electronic ratchets is extended to polarization-sensitive linear and circular ratchets. 
The ratchet currents which are sensitive to the linear and circular polarization of the light 
arise in the same system with broken symmetry due to the phase
shift between the periodic potential and the periodic light field resulting from near field diffraction. 
They appear because the carriers in the laterally modulated quantum wells 
move in two directions and are subjected to the action of the two-component electric field.
In contrast to the photothermal current, the linear and circular ratchet currents are independent of the energy relaxation time
and controlled only by the momentum relaxation time.

\acknowledgments The financial support from the DFG (SPP~1459, GK~1570) and RFBR is
gratefully acknowledged. We are grateful to M. Grifoni, M.~B\"uttiker 
and V.~V. Bel'kov for fruitful discussions
as well as C.~Linz and R.~Ravash for sample preparation.

\end{document}